# Water-based Reconfigurable Frequency Selective Rasorber with Thermally Tunable Absorption Band

**Xiangxi Yan[1], Xiangkun Kong[1,2], Qi Wang[1], Lei Xing[1], Feng Xue[1], Yan Xu[1], Shunliu Jiang[1]**

[1] Key Laboratory of Radar Imaging and Microwave Photonics, Ministry of Education, Nanjing University of Aeronautics and Astronautics, Nanjing 210016, People's Republic of China
[2] State Key Laboratory of Millimeter Waves of Southeast University, Nanjing 210096, People's Republic of China

E-mail: xkkong@nuaa.edu.cn

**Abstract**

In this paper, a novel water-based reconfigurable frequency selective rasorber (FSR) at microwave band is proposed, which has a thermally tunable absorption band above the transmission band. The water-based FSR consists of a bandpass type frequency selective surface (FSS) and a 3D printing container. The water substrate is filled into the sealed space constructed by the above two structures. The numerical simulation results show that the FSR can achieve absorption with high absorptivity from 8.3 to 15.2 GHz, and obtain a transmission band of 5.2 to 7.0 GHz. The minimum insertion loss of the transmission band reaches 0.72 dB at 6.14 GHz. In addition, the FSR has the reconfigurable characteristics of absorbing or reflecting electromagnetic waves by filling with water or not. The proposed water-based FSR shows its good transmission/absorption performance under different polarizations and oblique incident angles. Due to the Debye model of water, the absorption band can be adjusted by water temperature, while the passband remains stable. At last, prototype of the FSR based on water has been fabricated, and the experimental results are presented to demonstrate the validity of the proposed structure.

## 1. Introduction

Frequency selective surfaces (FSSs) have been found numerous applications due to their attractive frequency filtering characteristics [1]. FSSs can be regarded as a spatial filter which performing a bandpass/bandstop response for electromagnetic wave, and have been widely used as radome, reflectors, and substrates of antennas. The FSSs with bandpass performance are employed in a radome design to suppress mutual interference, enhance radiation performance or reduce the Radar Cross Section (RCS) [2-4]. But such radomes based on FSSs usually reflect the out-of-band signals and result in lager RCSs in other directions, so they can barely reduce the mono-static RCS. In order to realize the purpose of the true stealth radome system, metamaterial absorbers (MMAs) are usually applied to design new type of radomes which can maintain the transmission at the passband but also absorb the out-of-band waves to reduce the bistatic RCS. Such a low RCS radome based on FSS and MMA is named as frequency selective rasorber (FSR), where the word "rasorber" is a combination of the terms "radome" and "absorber" [5].

In fact, various designs of FSRs have been proposed in recent years. The reported FSRs can be mainly divided into 2-D FSR and 3-D FSR according to the structural dimensions. Traditionally, the 2-D FSRs are usually realized by cascading a lossy FSS with a lossless FSS [6-11]. The incident waves of frequencies outside the passband can be absorbed by lossy FSS similar to the operating principle of Salisbury Screens, while retaining the transmission performance at the passband frequencies. The 3-D FSRs are mainly constructed by parallel-plate waveguides (PPW) to realize multiple resonators or propagation modes. Based on this innovative concept, the 3-D FSRs can be designed to achieve high selectivity, low insertion loss or wide absorption band [12-15]. Of course, the existing FSRs can be sorted into three categories based on the relative locations of the transmission band and the absorption band: that with the transmission band above the absorption band [16-17], that with the transmission band below the absorption band [6],[12-13],[18], or the transmission band between two absorption bands [7],[9-11].

But those mentioned above FSRs both can be defined as passive FSRs, which the structures, once built, are generally inflexibility, because of their fixed nature after fabrication. Unfortunately, in practical applications, the requirement of tunability of FSRs is often necessary because of the change of the realistic environment, and traditional passive FSRs would be seriously limit its scope of applications. So the active FSRs (AFSRs) is proposed, which can realize the performance of controlling electromagnetic characteristics through actively adjusting external excitation (e.g., dc voltage size, field energy and so on). The reported electrically controlled AFSRs are divided into two categories: the switchable type of AFSRs [19-21] and the tunable type of AFSRs [18], [22]. In general, the AFSRs are commonly loaded with active components, such as PIN diodes and varactor diodes, to achieve the reconfigurable performance. However, many extra feeder lines and bias networks are employed in the structure array, causing design complexity, affecting the reconfigurable performance, high fabrication cost and difficult measured process. In order to simplify the design procedure of the AFSRs and also realize the reconfigurable purpose of rasorber structures, a novel FSR based on water substrate is firstly proposed in this paper. Water substrate is exactly applied to overcome the above disadvantages.

According to the particular physics characteristics, water has been found many applications in the design of water antennas [23-24] or filling in metamaterial absorbers as substrate. Traditionally, MMAs usually adopt loading lumped resistive [25-27], using high-impedance surfaces [28-30] or magnetic materials [31] to realize broadband absorption. Water as a unique liquid material can provide a novel theory for the design of broadband MMAs. For one thing, water has a high value about permittivity and dielectric loss in terms of electrical parameters due to the dispersion characteristic. Water can be used as a multi-layered substrate with the container for supporting multiple resonance modes, so water-based "all-dielectric" MMAs have ultra-broadband absorption operating frequency [32-37]. For another thing, as a liquid, water can be injected into various shapes of a container as a substrate due to fluidity. Compared with traditional MMAs, the absorbers based on water can highly simplify the design procedure, reduce fabrication cost and realize the optical transparency performance. As an example, Qu et al. developed a broadband absorption and optical transparency MMA based on water, polymethyl methacrylate (PMMA) and indium tin oxide (ITO) substrate [37]. The developed MMA achieved broadband microwave absorption with efficiency larger than 90% in 6.4 ~ 30 GHz as well as a high optical transmittance of about 85%. In addition, the thermal tunability performance of such all-dielectric MMAs can be realized by water temperature. So water-based structures have great potential value in the practical application. However, due to the information available to the author, all water-based MMA structures typically have a metal or ITO backing plate at the bottom, and no water-based FSR with a transmission band has been designed.

In this study, a new kind of FSR structure based on water substrate is firstly proposed. The water-based FSR composed of a bandpass type FSS and a 3D printing container. The water substrate is filled into the sealed space constructed by the above two structures. The proposed FSR can realize broadband absorption from 8.37 to 15 GHz, and obtain a transmission band covers 5.2 to 7.0 GHz. Furthermore, the transmission and absorption characteristics of the FSR under different polarizations and oblique incident angles is analysed, showing the stability of the passband and -10dB absorption band. The thermal tunability and reconfigurability of the proposed rasorber are also investigated. At last, a prototype of the proposed FSR structure is fabricated and experimentally tested. Good agreement with the simulated and the measured results indicate the validity of the proposed design.

## 2 Water-based FSR

The proposed water-based FSR is a sandwiched structure consisting of a container layer, a water layer, and an FSS layer, as shown in Fig.1. On the bottom layer, the FSS is made up of the cross-slot type unit that is printed on the $F_4B$ substrate. The thickness of the $F_4B$ substrate is $h_1$. The length and width of the cross-slot type unit are $l$ and $w$. As shown in Fig.1 (b), the upper layer structure is constructed of resin material by using 3D printing technique, whose shape is the same as that of the FSS layer. A sealed space between the two layers is designed to fill with water substrate. In Fig.1(c), the thickness of the container, the water layer, and the $F_4B$ substrate are $h_3$, $h_2$ and $h_1$, respectively. The geometrical parameters are $p$=20 mm, $l$=18 mm, $w$=3 mm, $h_1$=0.5 mm, $h_2$=0.9 mm, $h_3$=3.8 mm.

The S-Parameters of the water-based FSR are numerically investigated using the full-wave electromagnetic simulation software CST Microwave Studio. In the simulation, the permittivity of the resin and the $F_4B$ are 2.93(1-j0.04) and 2.65(1-j0.001), respectively. The permittivity of water at radio frequency can be described by the Debye formula with the filled water temperature of $T$ as follows [38]:

$$\varepsilon(\omega,T) = \varepsilon_\infty(T) + \frac{\varepsilon_0(T) - \varepsilon_\infty(T)}{1 - i\omega\tau(T)}, \quad (1)$$

where $\varepsilon_\infty(T)$, $\varepsilon_0(T)$ and $\tau(T)$ are the optical permittivity, static permittivity and rotational relaxation time, respectively. Meanwhile, $\varepsilon_\infty(T)$, $\varepsilon_0(T)$ and $\tau(T)$ in the Debye formula of water are all directly dependent on water temperature $T$:

$$\varepsilon_0(T) = a_1 - b_1 T + c_1 T^2 - d_1 T^3, \quad (2)$$

$$\varepsilon_\infty(T) = \varepsilon_0(T) - a_2 e^{-b_2 T}, \quad (3)$$

$$\tau(T) = c_2 e^{\frac{d_2}{T+T_0}}, \quad (4)$$

where $a_1$=87.9, $b_1$=0.404 $K^{-1}$, $c_1$=9.59×10$^{-4}$ $K^{-2}$, $d_1$=1.33×10$^{-6}$ $K^{-3}$, $a_2$=80.7, $b_2$=4.42×10$^{-3}$ $K^{-1}$, $c_2$=1.37×10$^{-13}$ $K^2$, $d_2$=651 ℃, $T_0$=133 ℃ and $T$ is the water temperature in ℃.

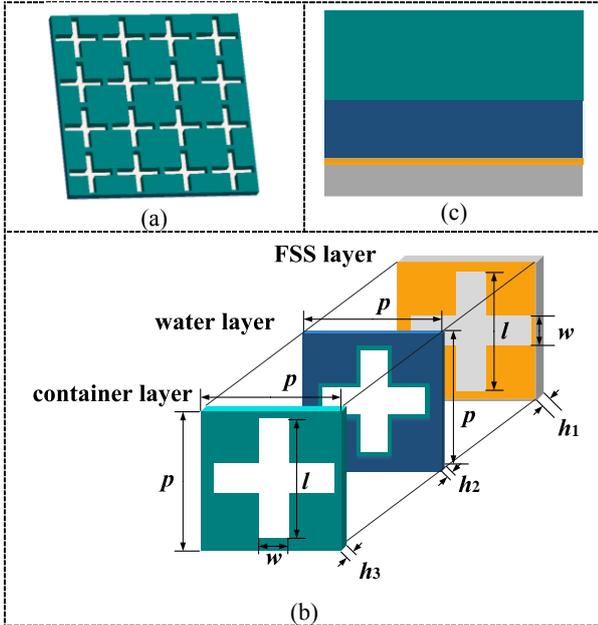

Fig 1. Schematic diagram of the water-based FSR. (a) 3D periodic structure; (b) layer by layer of the unit cell; (c) side view of the unit cell.

The real and imaginary parts of the permittivity change significantly at different water temperatures, as shown in Fig.2. On the one hand, the effective absorption in microwave frequency can be achieved due to the high imaginary part of the permittivity of water. On the other hand, the water-based FSR has the ability of thermal tunability owing to the significant dispersion characteristics of the permittivity of water.

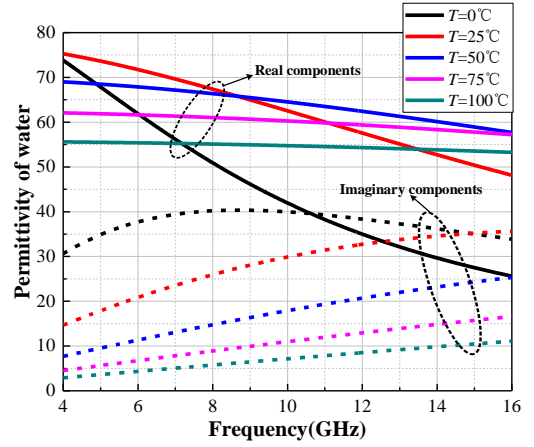

Fig 2. Dielectric permittivity of water for different temperature.

## 3 Results and performance analysis

### A Reconfigurable performance

Firstly, we simulate the model of water-based FSR at room temperature (25℃). The numerical simulation results of transmission/reflection coefficient under normal incidence are shown in Fig.3 (a). Because of the bandpass response of the FSS layer, the water-based FSR can obtain a transmission window with a minimum insertion loss of 0.72 dB between 5.2 GHz and 7.0 GHz. The absorption band is achieved from 8.3 GHz to 15.2 GHz with the absorptivity higher than 80%, where the $S_{11}$ and $S_{21}$ are less than -10 dB.

Secondly, the electromagnetic characteristics of the proposed structure without filled water are studied. The simulation results are shown in Fig 3 (b). It can be noticed that the previous absorption band in the high frequency range changes into the reflection band, while the transmission band does not shift significantly. The results demonstrate that the water-based FSR has good reconfigurable performance due to the filled water in the structure has absorption effect on electromagnetic wave. In Fig.3 (b), the gap between the lower and upper effective reflection band is caused by the dielectric resonance of the container substrate. There is a slight offset for the transmission zero of the passband because of the chang of the effective dielectric constant and equivalent permeability caused by filling with water or not. In summary, the designed FSR can realize the reconfigurable performance with or without filled water.

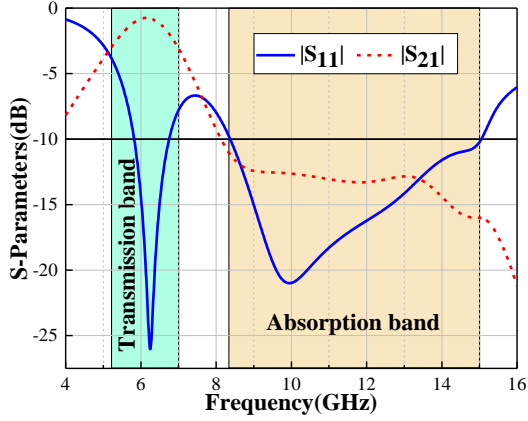

(a)

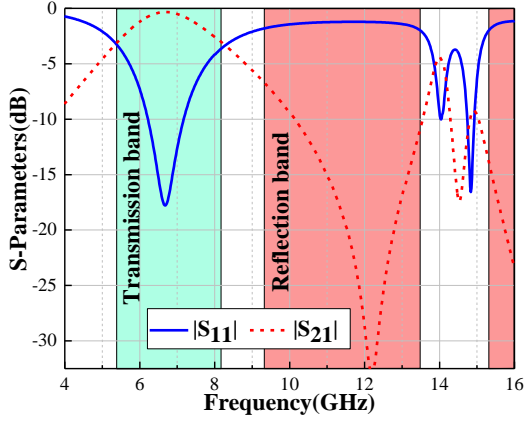

(b)

Fig 3. Reconfigurable performance of the water-based FSR for different states. (a) with water; (b) without water.

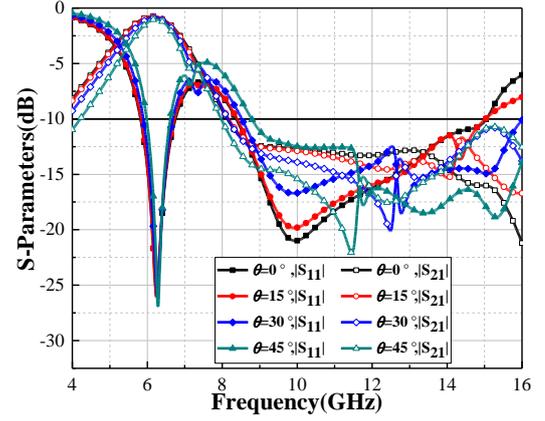

(a)

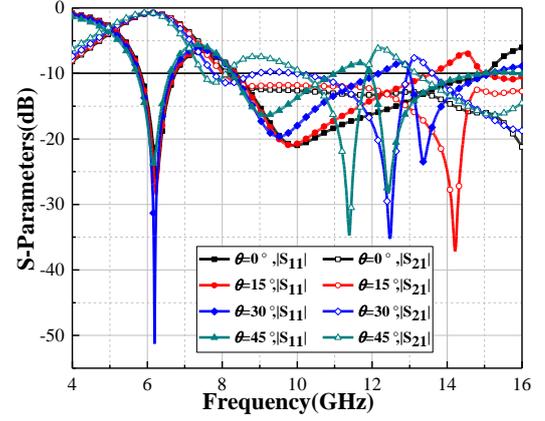

(b)

Fig 4. Simulated transmission/reflection coefficients of the water-based FSR with different incident angles. (a) TE polarization; (b) TM polarization

## B Wide-angle performance

Fig 4 shows the transmission/reflection coefficients of the water-based FSR under oblique incidence of electromagnetic waves. As shown in Fig. 4(a), for TE model, when the oblique incident angle increases from 0° to 45°, the intensity of the magnetic field decreases gradually and the electric field remains unchanged, which makes the effect of incident angle on dielectric resonance weaker. The absorptivity is still higher than 80% in the entire absorption band under wide-angle incidences. As shown in Fig.4 (b), for TM model, the electric field intensity decreases significantly with the increase of incident angle, and the dielectric resonant is sensitive to the electric field. As a result, the absorption band decreases remarkably when the incident angle is less than 45°, except for several special frequency points, but the absorption rate maintains high. In the meantime, with the increase of incident angle, the absorption band shifts to higher frequency. Furthermore, the transmission window under both TE and TM polarization waves exhibits the same bandpass response. Under all these conditions, the transmission zero of the passband is consistently located at 6.24 GHz. Based on the above analysis, the proposed water-based FSR can display good bipolarization and wide incident angle performance.

## C Thermally tunable performance

As mentioned in Fig.2, the permittivity of water has a certain relationship with its temperature $T$, so it is necessary to study the thermal tunability of water-based rasorbers. According to the formulas (1) to (4), the numerical simulation results are obtained by changing the water temperature, inwhich only the TE wave under normal incidence is considered. Fig.5 (a) presents the $S_{11}$ and $S_{21}$ of the transmission window for the temperature varies from 0~100℃. It can be seen that the transmission window maintains the constant bandwidth of the -3 dB passband and has the function of thermal stability. The transmission window is influenced weakly at different water temperatures, because the permittivity of the substrate of whole structure has little change. In contrast, as shown in Fig.5 (b), the absorption band exhibits good thermal tunability with the increase of the water temperature: in the lower frequency, the absorption bandwidth gradually becomes narrow and the fractional bandwidth obviously decreases, while in the higher frequency, the variation of the absorption bandwidth is very small. Meanwhile, when the temperature increases, the absorptivity falls slowly, but it still keeps over 80%. Therefore, the water-based FSR exhibits the thermal stability of the transmission window and the active thermal adjustment of the absorption band which varies with the water temperature.

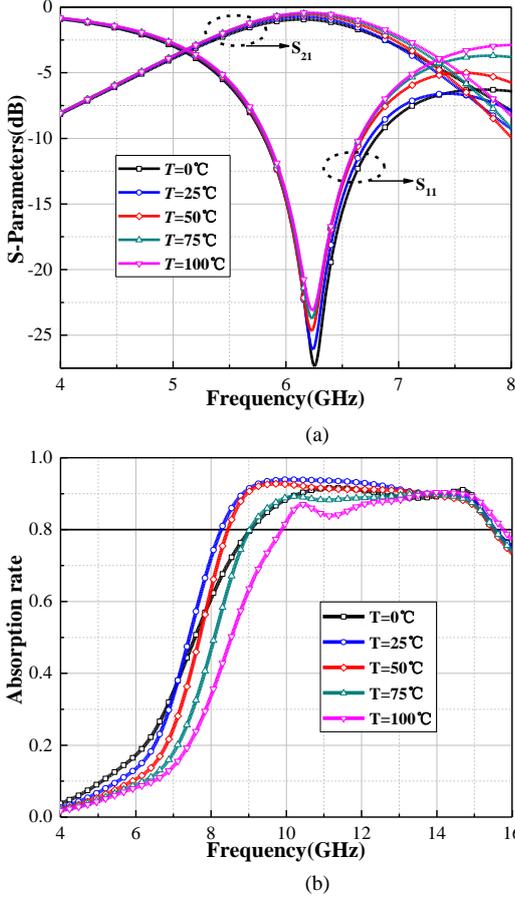

Fig 5 The transmission/absorption results of the water-based FSR under different temperatures. (a) Transmission window; (b) Absorption band.

*D Mie resonance theory analysis based on water*

Moreover, we consider the transmission/absorption band under the condition of optimizing structure parameters. Fig 6 presents the S-Parameters results of the water-based FSR with different thickness of water layer $h_2$. For the absorption band, when the thickness of $h_2$ varies from 0.8 to 1.0mm, the relative bandwidth of the absorption band can be broadened gradually, and the absorptivity keeps higher than 80% at all times. For the transmission band, the bandwidth of -3 dB passband and the minimum insertion loss almost remain unchanged with the thickness of $h_2$ increase to 0.9 mm. When $h_2$ up to 1mm, the minimum insertion loss of the transmission band reach to 1 dB, which is worse than that of the above situation. Thus, in pursuit of the minimum insertion loss of the passband and expand the absorption bandwidth simultaneously, we choose the appropriate thickness $h_2$=0.9mm as the optimal structural parameter of the aquifer.

It can be seen that when the value of $h_2$ increases, the resonance frequency moves to lower frequency. The reason is that under the excitation of electromagnetic wave, a mixed dielectric layer consisting of water and resin substrate can produce dielectric resonance in the proposed composite structure. The resonance frequency can be studied by analyzing the Mie resonance theory [39] as:

$$f = \theta c \big/ 2\pi r \sqrt{\varepsilon_p \mu_p} , \quad (5)$$

$$F(\theta) = \frac{2(\sin\theta - \theta\cos\theta)}{(\theta^2 - 1)\sin\theta + \theta\cos\theta} , \quad (6)$$

where $F(\theta)$ is a resonant function and becomes infinite at some values of $\theta$, so that results in the magnetic and electric resonances. Among all the resonances, the first-order Mie resonance mode is the most intense. $\theta$ is a function about the resonance frequency and approximately equally to $\pi$ for the first-order Mie resonance; $c$ is the speed of light in vacuum and $r$ is the radius of the dielectric sphere. $\varepsilon_p$ and $u_p$ are the permittivity and permeability of the dielectric particles. However, it should be noted that in article [40-41], the dielectric materials used to calculate the first Mie resonance are usually cubic or sphere particles, but here, the water in container forms a rectangle, so it is no longer applicable to calculate the resonance frequency by using the parameter $r$ in equation (5). Therefore, we provide a modified formula (7) to illustrate the specific relationship between $h_2$ and $r$, and further get the functional relationship between $h_2$ and the resonance frequency. The corrected formula can be written as

$$r = \pi^2 \times \left(\frac{3}{4}\right)^{5(h_2 - 1)} , \quad (7)$$

In order to verify the validity of the mordified formula, the results calculated by Eq. (7) are compared with those obtained by simulation software, as shown in Fig 7. The results show the theoretical calculation results are in good agreement with the simulation results, which proves the correctness of the above formulas.

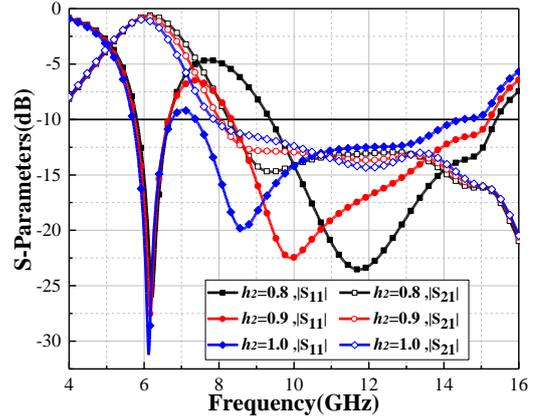

Fig 6 Simulated transmission/reflection coefficients by varying the thickness $h_2$ of the water layer.

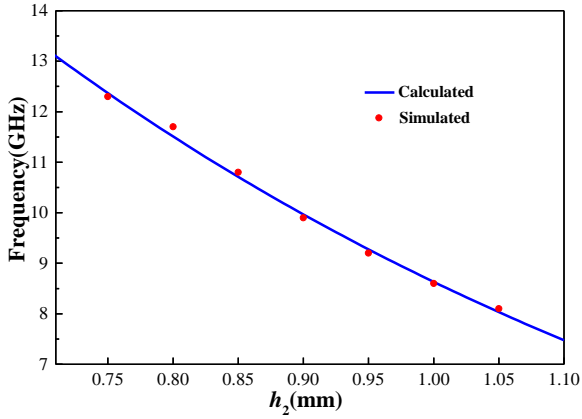

Fig 7 Comparison of the simulated and theoretical resonance frequencies with different thickness of $h_2$.

## 4 Experimental Verification and Discussion

A prototype of the proposed rasorber is fabricated to validate the main functions, as shown in Fig.8. The water-based FSR is composed of 15 × 15 unit cells with the overall size of 300 mm × 300 mm × 5.2 mm. The upper layer plays a role of container constructed by using 3D printing technology. The bottom layer is made up of the bandpass FSS, which is printed on $F_4BK265$ substrate. The epoxy resin adhesive with 0.5mm thick is used to bond the upper and bottom layer to build a sealed space for filling with water substrate. Based on the theory of connected container, two holes of connected hose are arranged to fill and drain water.

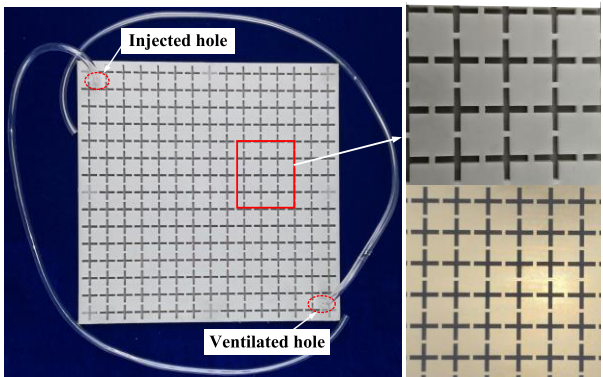

Fig 8 prototype of the water-based FSR.

Fig. 9 shows the measurement environment of transmission /reflection performance of water-based FSR prototype in anechoic chamber. The prototype is set in the middle of the measurement platform. Two standard horn antennas (2-18 GHz), mounted on both ends of the measurement platform for aiming at the center of the prototype, are connected to an Agilent N5245A vector network analyzer (VNA) by using two RF cables. The measured results of the prototype with and without filled water under the normal incidence at room temperature are shown in Fig.10 (a) and (b). Considering the effcet of the exoxy resin adhesive between the two layers, we optimize the previous model and show the optimized simulation results of the water-based FSR with and without filled water in Fig.10. The comparison of the former and optimized simulation results is shown in Fig.3 (a) and Fig.10 (a), it can be seen that the optimized transmission band shifts to lower frequency, while the absorption band stays the same. In Figs.10 (a) and (b), the measured results have good agreement with the optimized ones. Moreover, there is a slight deviation caused by the fabrication tolerance of 3D technology.

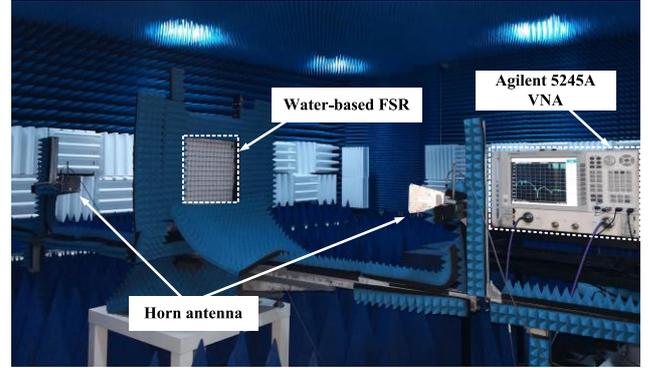

Fig 9 Measurement setup.

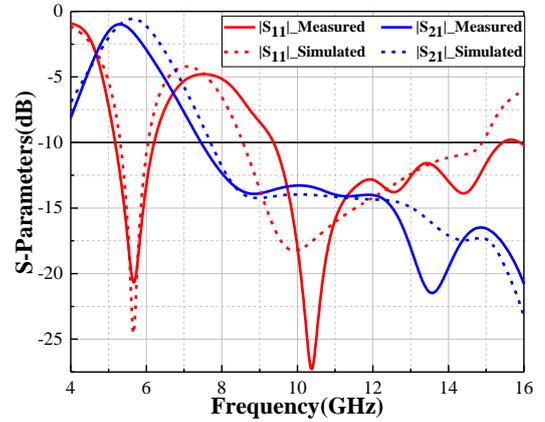

(a)

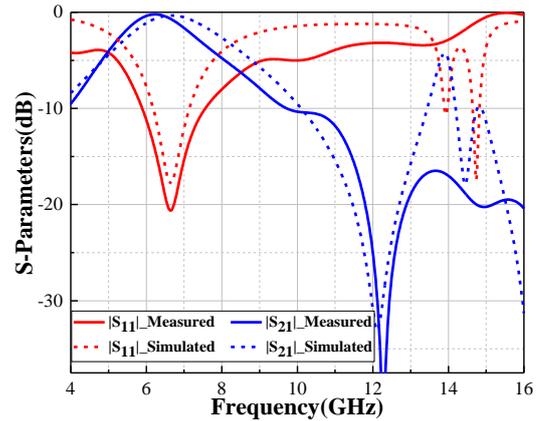

(b)

Fig 10 Comparison of measured and re-simulated results of the water-based FSR under normal incidence. (a) with water; (b) without water.

Next, the wide-angle performance of our designed water-based FSR is further studied, as shown in Fig11. The measured results of water-based FSR under both TE and TM polarized wave at incident angles of 15 ° and 45 ° are in good agreement with the optimized results. It is noticed that there are two harmonics in the simulation results, which is due to the existence of harmonics of the bandpass FSS at its rejection band.

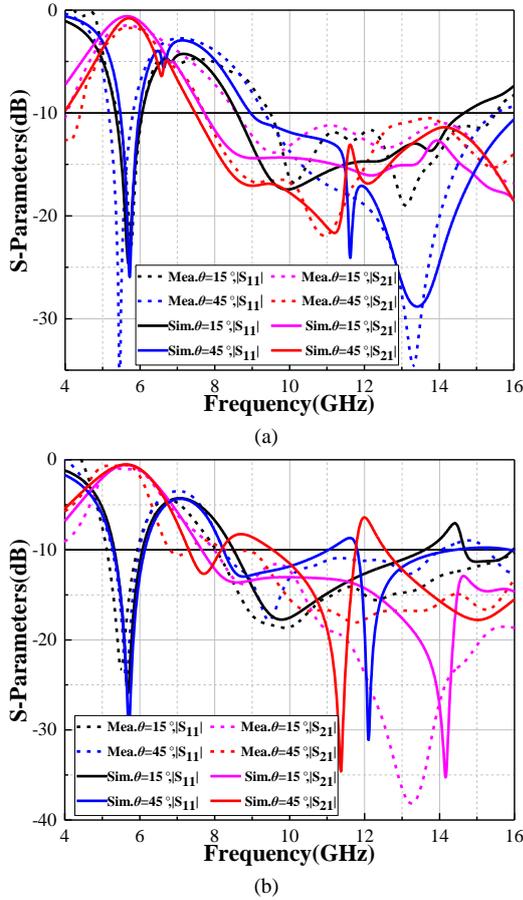

Fig 11 Measured and re-simulated transmission/reflection coefficient of the water-based FSR under oblique incidence. (a) TE polarization; (b) TM polarization.

Finally, the transmission/reflection coefficients of water-based FSR at different temperatures are discussed. A thermostat is introduced to create a same measurement environment as the simulation ones. Owing to the volatility of the epoxy resin adhesive and the thermal deformation of the resin material, the measured temperatures are set below 55℃, and the transmission/reflection coefficients at different temperatures are measured, as shown in Fig.12. The measured transmission zero of the passband slightly shifts to a lower frequency when the temperature goes up from 40℃ to 55℃. The discrepancies of the measured and optimized results are generated by the thermal influence of various structural materials and the edge scattering caused by processing size, which are not considered in our model simulation. From the above analysis, the measured results effectively prove the feasibility of the proposed water-based rasober.

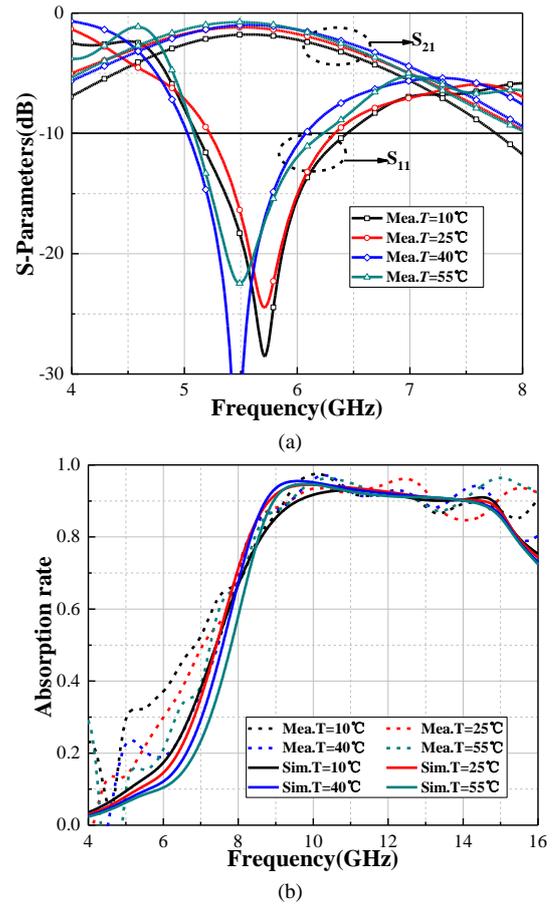

Fig 12 Measured and re-simulated results of the water-based FSR at different temperature. (a) transimission window; (b) absorption band.

## 5 Conclusion

In conclusion, a water-based reconfigurable FSR is designed to achieve a thermally tunable absorption band while ensuring a high selective transmission band. We adopt FSS layer and 3D printing container to assemble the proposed rasorber. The water layer can be created by filling with water into the sealed space, which is built by the upper container and the bottom FSS layer. The simulated -10 dB absorption band ranges from 8.3 to 15.2 GHz, and the transmission window is generated by the bandpass FSS in the frequency from 5.2 to 7.0 GHz. In addition, compared with traditional AFSRs, the reconfigurability of water-based FSR is obtained by filling with water or not. The wide-angle performance of FSR under both TE and TM polarizations waves is further analyzed and demonstrated. Moreover, the thermal tunability of water-based FSR at different temperatures is discussed. The structure is fabricated and the main functions obtained from theoretical analysis are verified by experiments. The innovative water-based rasorbers have potential application in antenna stealth technology.

**Acknowledgements**

The authors greatly appreciate the help of Professor Qiang Cheng from Southeast University for his meaningful advice and assistance in this paper. This work is supported by 'the Fundamental Research Funds for the Central Universities' (No. kfjj20180401),the Chinese Natural Science Foundation (Grant No.61471368), Aeronautical Science Foundation of China (20161852016), China Postdoctoral Science Foundation (Grant No. 2016M601802), Open Research Program in China's State Key Laboratory ofMillimeter Waves (grant No. K202027) and by Jiangsu Planned Projects for Postdoctoral Research Funds (Grant No. 1601009B).